\documentclass[preprint,pra,superscriptaddress]{revtex4}

\usepackage{epsfig}
\usepackage{graphicx}
\usepackage{amsmath}
\usepackage{amssymb}
\usepackage{bm}

\newcommand{\Xs}{{X^1\Sigma_g^+}}
\newcommand{\As}{{A^1\Sigma_u}}
\newcommand{\At}{{b^3\Pi_u}}

\newcommand{\W}{{\Omega}}
\newcommand{\G}{{\Gamma}}

\begin{document}
\title{Photoassociation adiabatic passage of ultracold
Rb atoms to form ultracold Rb$_2$ molecules}

\author{Evgeny A. Shapiro}
\affiliation{Department of Chemistry, University of British
Columbia, Vancouver, Canada}
\author{Moshe Shapiro}
\affiliation{Department of Chemistry, University of British
Columbia, Vancouver, Canada}

\affiliation{Department of Chemical Physics, Weizmann Institute of
Science, Rehovot 76100, Israel}
\author{Avi Pe'er}
\affiliation{JILA, National Institute of Standards and Technology
and University of Colorado, Boulder, CO 80309-0440}
\author{Jun Ye}
\affiliation{JILA, National Institute of Standards and Technology
and University of Colorado, Boulder, CO 80309-0440}

\begin{abstract}
We theoretically explore photoassociation by Adiabatic Passage of
two colliding cold $^{85}$Rb atoms in an atomic trap to form an
ultracold Rb$_2$ molecule.
We consider the incoherent thermal nature of the scattering
process in a trap and show that coherent manipulations of the
atomic ensemble, such as adiabatic passage, are feasible if
performed within the coherence time window dictated by the
temperature, which is relatively long for cold atoms. We show that
a sequence of $\sim 2\times10^7$ pulses of moderate intensities,
each lasting $\sim750$ ns, can photoassociate a large fraction of
the atomic ensemble at temperature of $100$ $\mu$K and density of
$10^{11}$ atoms/cm$^3$. Use of multiple pulse sequences makes it
possible to populate the ground vibrational state. Employing
spontaneous decay from a {selected} excited state, one can
{accumulate the molecules in a narrow distribution of vibrational
states in the ground electronic potential}. Alternatively, by
removing the created molecules from the beam path between pulse
sets, one can create a low-density ensemble of molecules in their
ground ro-vibrational state.
\end{abstract}


\maketitle

\section{Introduction.}

The use of Adiabatic Passage (AP) \cite{stirap1,stirap2} to form
ultracold molecules by photoassociating ultracold atoms
\cite{vardi97,mackie-PRA99,stirap-arguments,SBbook,BECs} with
Resonantly Enhanced Anti-Stokes Raman (REAR) pulses in
counter-intuitive ordering is an attractive idea, mainly due to
the promise of high yield. Central to the AP process is the
formation of a ``dark state'' i.e., a light-dressed state that
once formed, is impervious to further actions of the light fields.
When the light fields are applied as pulses, AP results from the
dark state changing its nature by following the makeup of the
applied pulses from that of the input state (scattering state) to
the desired final state (molecular bound state).

Unlike in conventional bound-bound AP (e.g. STIRAP)
 \cite{stirap1,stirap2}, the input state in the photoassociation is
not a pure state, but a thermal mixture of states. It can be
represented as an incoherent mixture of different energy
eigenstates, where the scenario of coherent manipulation proposed
here works with similar efficiency for any initial energy within
the thermal spread of the initial ensemble. Alternatively, the
input state can be represented as a mixture Gaussian wave packets
localized in phase space. The applied laser pulses in our scheme
are not much longer than the coherence time of the thermal
ensemble, which is equivalent to the average collision time of the
coherent wave packets. Within the coherence time-window all
initial states behave alike, and the coherent manipulation with
these states succeeds.

Another difference between AP from the continuum and conventional
three-level STIRAP due to the dynamical nature of the continuum is
that while in STIRAP the input state can be completely transferred
into the final state, in photoassociation this is not possible
(only a small fraction of the atoms are colliding at a given
time). Here however, the continuum serves as a coherent source /
sink of population that is coupled by the light fields to the
bound molecular states. By applying the coherent pulses over and
over on the thermal ensemble of atoms, the overall
photoassociation yield, which is an incoherent sum of yields from
individual pulse sets, can approach unity on a reasonable time
scale.

Although dark states connecting continuum and bound states have
now been observed experimentally
 \cite{lett-darkstate-PRA05,winkler-BECdarkstate-prl05},
Photoassociation Adiabatic Passage (PAP) has not been realized so
far. The main reason is that adiabaticity implies large
``Rabi-frequencies'', which, due to the weakness of the bound
continuum dipole-matrix-elements (controlled by the Franck-Condon
(FC) shape factors), imply the use of high intensity pulses. The
production of high intensity in conjunction with relatively long
pulse durations is not so easy to achieve in practice. In
addition, the exact values of the pulse parameters are still under
study  \cite{stirap-arguments}.

This paper describes a set of realistic calculations of the
$^{85}$Rb+$^{85}$Rb $\rightarrow$ $^{85}$Rb$_2$ photoassociation
in an attempt to determine the pulse parameters needed to bring
about PAP for this system. We use two pairs of pulses: The first
of these pairs transfers the population from continuum to the
fourth excited state of the $X^1\Sigma_g^+$ of Rb$_2$ (REAR). The
second pair transfers the population to the ground vibrational
state (conventional STIRAP). Special care is taken of averaging
over the thermal ensemble of colliding atoms at 100 $\mu$K
temperature, employing the laser pulses that photoassociate most
initial states in the ensemble.
We find that with sensible laser parameters {photoassociation at a
rate of $\sim5\times10^{-8}$ per pulse is possible, which reflects
the limit set by the sample density and temperature. Accordingly,
photoassociation of a major fraction of the ensemble can be
achieved by repeating the pulse sequence $\sim 2\times10^7$ times.
In order to avoid excitation of the created molecules by
subsequent pulses it is necessary to either remove them from the
beam path or ``hide'' them via an irreversible process (e.g.
spontaneous emission). { We discuss both the possibility of using
a slowly moving optical lattice to remove the molecules form the
laser focus, and a method involving {spontaneous} decay from an
excited molecular state.} In the case of photoassociation of high
phase-space density sample, it must be possible to photo-associate
the whole sample with a single pulse sequence.}

\section{Molecular data.}

\begin{figure}
\centering
    \includegraphics[width=0.65\columnwidth]{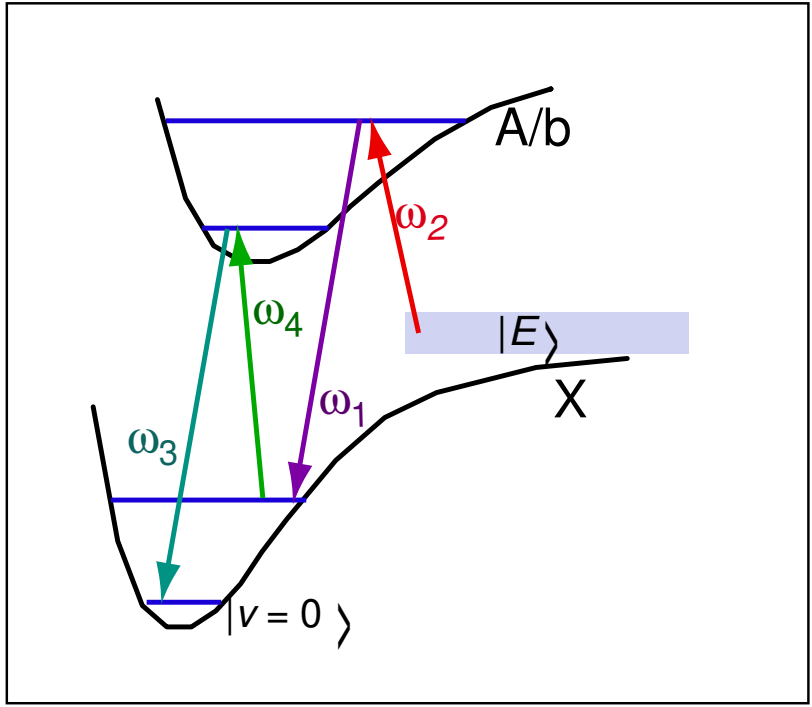}
    \caption{(Color online). The combination of
    photoassociation adiabatic passage (PAP) pairs discussed in the text.}
 \label{FigThreeways}\end{figure}

For notational simplicity we henceforth denote the $\Xs$
electronic state as ``X'', and the spin-orbit-coupled $\At$ and
$\As$ potentials as ``A/b''. We shall consider the
photoassociation scenario illustrated in Fig.~\ref{FigThreeways},
starting with two $^{85}$Rb atoms colliding on the Rb$_2$
X-potential. A pair of laser pulses transfers, using Resonantly
Enhanced Anti-stokes Raman (REAR), a fraction of the incoming
continuum population to an excited vibrational state of the Rb$_2$
X-state. The pair of pulses consists of a pump pulse, which
couples the X-continuum to one of the A/b-bound states, and an
Anti-Stokes dump pulse, which couples this A/b-bound state to an
excited X-vibrational state. A second, much weaker, pair of
pulses, can transfers (again by REAR, which in this case reduces
to the bound-bound STIRAP \cite{stirap1,stirap2})
 the population from the excited X-vibrational state to the ground
vibrational state. {Alternatively, the population can be moved to
the ground state through {spontaneous} decay of an excited
molecular level.}

\begin{figure}
\centering
    \includegraphics[width=0.87\columnwidth]{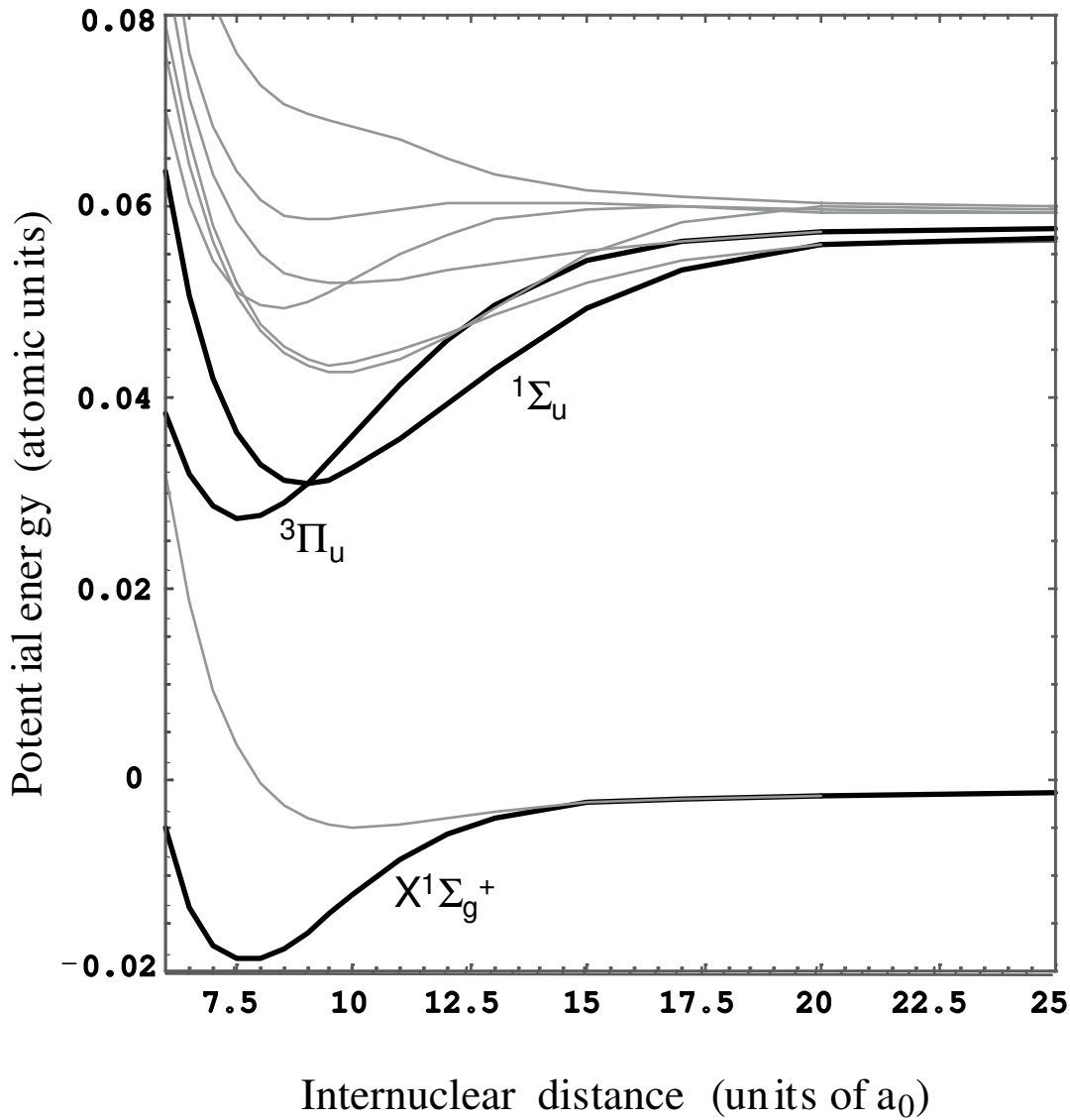}
     \caption{Black lines: Rb$_2$ Born-Oppenheimer potentials involved in the
photoassociation calculations.
     Gray lines: Rb$_2$ potentials not used in this paper.
The data shown in the Figure are adopted from Ref.
\cite{spiegelmann-JPB89}.}
     \label{FigPotentials}
\end{figure}
\begin{figure}
\centering
    \includegraphics[width=0.8\columnwidth]{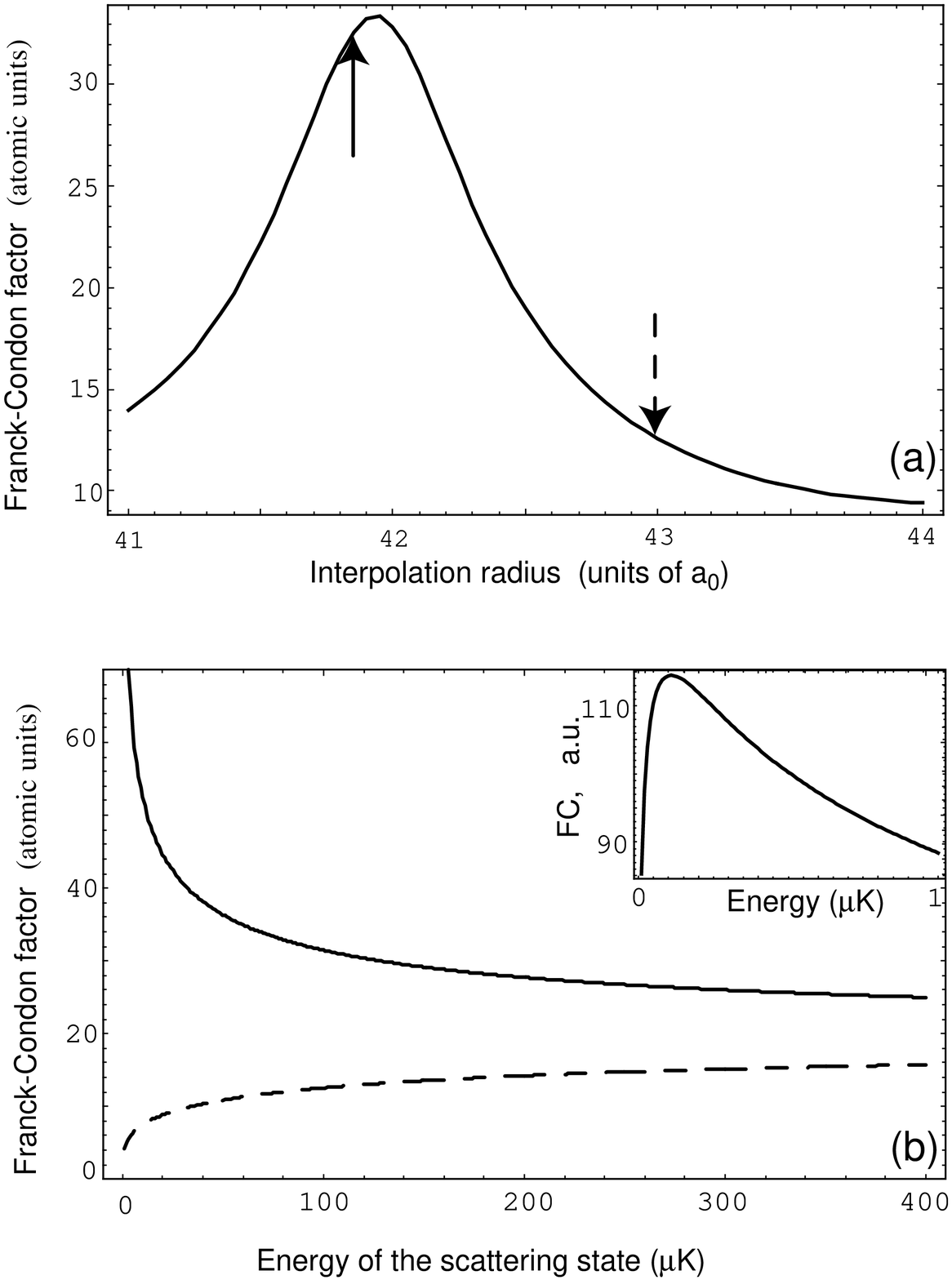}
    \caption{(a) The Franck-Condon (FC) factors for the
transition from a continuum state of 100 $\mu$K energy to the
A/b(133) level, for different X-potential long-range/short-range
interpolation radii. The full arrow indicates the fitting
parameter that correctly yields the measured scattering resonance
and scattering length, which is used in further calculations.
(b) The continuum-bound FC factors for the transition to the A/b$(v=133,J=1)$ level
as a function of the scattering energy. The full (dashed) line corresponds
to an interpolation radius marked by the full (dashed) arrow in (a).
The inset shows the FC factor near the resonance
at the $E=0-1$ $\mu$K continuum energy range.}
     \label{FigFCfits}
\end{figure}

Fig.~\ref{FigPotentials} shows the Born-Oppenheimer potentials
used in the calculations. The X-potential at inter-atomic
distances below 40 a.u. was taken from the Ref.
 \cite{edvadsson-03}, whereas at large distances the X-potential
was modelled as a $-C_6/r^6$ dispersion term, with the value $C_6=
4426$ a.u. taken from Ref.~ \cite{marinescu-PRA94}. The short-term
and the long-term potentials were smoothly interpolated near 40
a.u. The A/b potentials, and the spin-orbit coupling between them
were adopted from Ref. \cite{edvadsson-03}.

The continuum-bound FC factors for the X-A/b transitions were
calculated using the artificial channel method
 \cite{moshe-acm-72}.
Their scaling was verified by comparison with that of the s-wave
scattering wavefunction found by a simple outward propagation on
the X-potential. Unlike bound-bound FCs, which are dimensionless
overlap integrals, the free-bound FCs have dimensions of
$(Energy)^{-1/2}$, due to the different normalization of free
states.

The s-wave scattering of the two $^{85}$Rb atoms on the
X-potential at the energies of interest is influenced by a
resonance, giving rise to a scattering length in excess of 2400
a.u.  \cite{scl}. This resonance is due to the last (quasi-) bound
state lying just a notch (above) below the continuum
 \cite{LL3-book,sc-reviews}. We mention in passing that the
collision of two $^{85}$Rb atoms on the $1^3\Sigma_u$ potential is
also influenced by a resonance, but to a lesser degree (scattering
length of $\sim360$ a.u). In contrast, the collision of two
$^{87}$Rb atoms at the energy range below is not influenced by the
resonance because it is sufficiently removed from it \cite{scl}.

The existence of a scattering resonance affects our
photoassociation scheme in two ways: First, the amplitude of the
continuum wavefunction in the inner region is strongly enhanced
relative to the non-resonant case. This enhances the X-A/b
continuum-bound FC factors, with the required laser intensities
being much lower than those needed in a non-resonant case. Second,
at near-resonance energies the $T(E)\sim E^{1/4}$ threshold law
\cite{sc-reviews} does not hold. As a result the FC factors at
very low energies are sufficiently high to ensure complete
adiabatic transfer, with the PAP scheme expected to work for
relatively low laser intensities.

The resonance features and the continuum-bound FC factors depend
sensitively on the way the interpolation between the short range
and long range forms of the X-potential is implemented.
Fig.~\ref{FigFCfits}(a) shows the FC factor to the
A/b$(v=133,J=1)$ ($E=0.042848$ a.u.) bound state for initial
scattering energy of 100 $\mu$K as a function of the X-potential
interpolation radius. If the latter quantity is chosen around 42
a.u. (full arrow in Fig. \ref{FigFCfits}(a)) the resulting
scattering length is $\sim2500$ a.u. When the interpolation radius
is chosen as 43 a.u. (dashed arrow) the resonance moves slightly
but sufficiently to change the scattering length to 100 a.u. The
dependence of the FC factors on the collision energy is shown in
Fig.~\ref{FigFCfits}(b). Away from the resonance the FC factors
follow the Wigner law.
The sensitivity of the FC factors to the interpolation radius
diminishes at higher energies.

\begin{figure}
\centering
    \includegraphics[width=0.8\columnwidth]{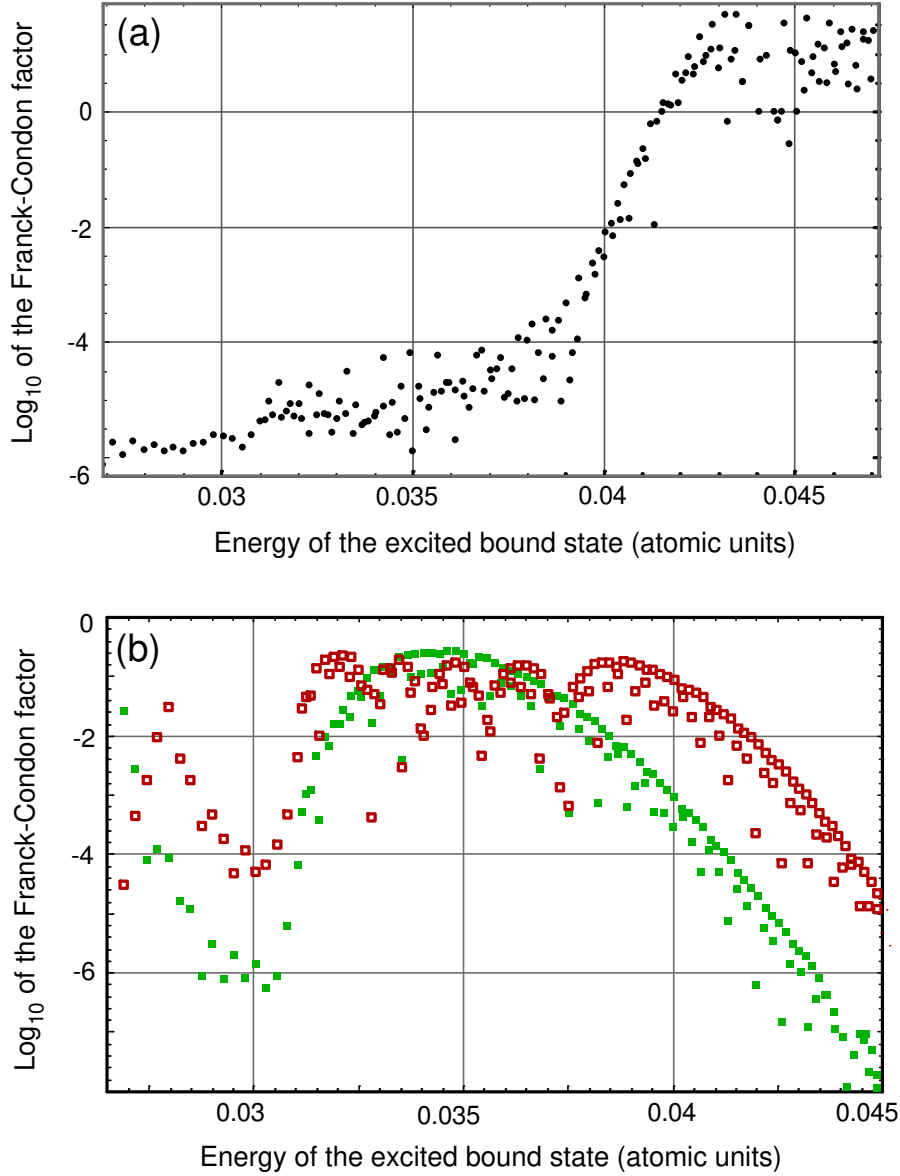}
    \caption{(Color online). (a) FC factors for the X$-$A/b continuum-bound
transitions. (b) FC factors for the A/b$-$X bound-bound
transitions, for the $v=0$ (green filled squares) and $v=4$ (red
open squares) X-vibrational states.}
     \label{FCs}
\end{figure}

The dependence of the continuum-bound s-wave FC factors for
scattering energy of $E=100$ $\mu K$ on the bound A/b states
energies is shown in Figure 4(a). The results can be explained using
a semiclassical phase space analysis based on the Husimi or Wigner
functions. This analysis shows that for the low-lying A/b states the
classical closed phase space trajectories lie at low position and
momentum values and do not intersect the X-state open collision
trajectory, rendering the A/b-X FC factor exponentially small. In
contrast, as the energy of the A/b states is increased, the X and
A/b phase space trajectories start to overlap, thereby increasing
the continuum-bound FC factors to several a.u.

The PAP scheme is expected to work conveniently for transitions to
bound states lying in the vicinity of the A/b$(v=133,J=1)$ level
($E=0.042848$ a.u.). This choice is based on the ability to
generate large area (microsecond-long) pulses at wavelengths near
$\sim1064$nm, corresponding to the transition frequency to this
state. The FC factor for the continuum-bound X$-$A/b$(v=133,J=1)$
transition, with initial scattering energy of 100 $\mu$K is equal
to 31.5 a.u.

Another scenario
 \cite{PAexps1,masnou1,masnou2,learningPAexp,chirpedPAexp} consists
of using the set of the highly extended ``sub-continuum'' bound
states existing just below the onset of the A/b continuum. The
advantage here is that the continuum-bound FC factors associated
with these states are higher by two or three orders in magnitude
relative to the A/b$(v=133,J=1)$ region, allowing for the use of
much weaker pump lasers. The disadvantage is that this scenario
necessitates controlling the long-range behavior of the
``sub-continuum'' states, which in turn depends very strongly on
the (poorly known) couplings between the many potentials which
play a part in the dynamics. Moreover, the ``sub-continuum''
states greatly weaken the A/b$-$X FC factors associated with the
dump pulse. Recent attempts to optimize photoassociation by
controlling behavior of these states have not brought a clear
answer on whether a significant optimization is possible
\cite{learningPAexp,chirpedPAexp}.

The eigenenergies of the coupled A/b bound states have been
computed by the artificial channel method \cite{moshe-acm-72} and
confirmed using the finite difference with Richardson
extrapolation FDEXTR code \cite{fdextr}. The latter was used to
compute the bound-bound X-A/b FC factors shown in
Fig.~\ref{FCs}(b).

In all the calculations reported below we estimate the X$-$A
electronic-dipole moment to be $\mu=3$ a.u. This value corresponds
to the $5S_{1/2}(m=1/2)-5P_{3/2}(m=3/2)$ transition in Rb atom in
a circularly polarized field. It is consistent with all the other
$5S-5P$ matrix elements for Rb atom under the action of a
polarized laser field. The exact values of the atomic reduced
dipole matrix elements can be found in Ref.~
\cite{safronova-alcalis-PRA99}.

\section{Theory of photoassociation adiabatic passage}

Following Ref.  \cite{moshe-disociation-JCP94,vardi97}, we express
the total Hamiltonian of our system as
\begin{equation}\label{hamiltonian}
  \hat H=\hat H_0 - 2 \hat\mu\sum_{n=1}^2\epsilon_n(t)\cos\omega_nt ,
\end{equation}
where $\hat H_0$ is the material Hamiltonian, $2\epsilon_n(t)$ are
the (slowly varying) amplitudes of the coupling fields and
$\hat\mu$ is the A/b$-$X electronic transition dipole moment. We
expand the material wavefunction as
\begin{equation}\label{bigexpansion}
\psi = \sum_{i=1}^2b_i e^{-iE_i t}|i\rangle + \int_{E_{th}}^\infty
dE \,b_E e^{-iEt}|E\rangle ,
\end{equation}
where $|i\rangle$ and $|E\rangle$ are bound and continuum states
of $\hat H_0$, $E_{th}$ is the threshold energy for the continuum
states,
\begin{equation}
(E_i-\hat{H_0})|i\rangle=(E-\hat{H_0})|E\rangle=0~,
\end{equation}
$i=1$ ($i=2$ ) denotes specific vibrational states in the X (A/b)
electronic manifold, and $|E\rangle$ denotes a continuum state in
the X electronic manifold. Atomic units ($\hbar=1$) are used
throughout.

We assume that the frequency $\omega_1$ of the laser field is in
near resonance with $\omega_{2,1}\equiv E_2-E_1$ (see Fig.
\ref{FigThreeways}) and that $\omega_2$ is in near resonance with
$\omega_{2,E}\equiv E_2-E$, thus justifying the use of the
Rotating Wave Approximation. The Schr\"odinger equation now
assumes the simple form  \cite{moshe-disociation-JCP94,vardi97}
\begin{eqnarray}
  \dot{b_{1}}&=& i \,\Omega_1^*\, b_2 e^{-i\,\Delta_1\,t}
~~~~~~~~~~~~~~~~~~~~~~~~~~~~~~~~~\label{RWA1-1}  \\
  \dot{b_{2}}&=& i\, \Omega_1 \,b_1\, e^{i\,\Delta_1\,t}
- \G_f b_2~ + i\,\int_{E_{th}}^\infty  dE \,
   b_{E}\, \Omega_E \,  e^{i\,\Delta_2\,t}
~\label{RWA1-2}\\
\dot{b_{E}}&=& i
   b_2 \,\Omega_E^*\, e^{-i\,\Delta_2\,t}
~~~~~~~~~~~~~~~~~~~~~\label{RWA1-3}
\end{eqnarray}
where $\Delta_1 \equiv \omega_{2,1}-\omega_1$, $\Delta_2 \equiv
\omega_{2,E}-\omega_2$ are the detunings and $\Omega_1 =
\epsilon_1\mu_{21}$, $\Omega_E = \epsilon_2\mu_{2E}$ are the Rabi
frequencies. The empirical term $\G_f b_2$ describes the free
decay of the $|2\rangle$ state. Since this free decay is
predominantly to high energy continuum states (hot, untrapped
atoms), it is assumed not to affect the continuum states
$|E\rangle$.

{\textrm Equations (\ref{RWA1-1}-\ref{RWA1-3}) can be simplified
in the following standard way  \cite{SBbook,vardi97}. Integrating
Eq.(\ref{RWA1-3}) gives
\begin{equation}\label{bE}
 b_{E}(t) =i\int_{0}^t dt'\,b_2(t') \,\Omega_E^*(t')\,
 e^{-i\,\Delta_2\,t'} + b_E(t=0),
\end{equation}
where the moment $t=0$ precedes the pulses. Upon substitution of
this equation into Eq.(\ref{RWA1-2}) one obtains
\begin{equation}
\dot{b_{2}}   
 = i\, \Omega_1 \,b_1\, e^{i\,\Delta_1\,t} - \G_f b_2~
  - \epsilon_2(t) \int_{0}^t dt'\,F_{corr}(t-t')\,\epsilon_2^*(t')\,b_2(t')
+ i\, \Omega_E\,F_0(t) %
\label{b2}\end{equation}
 Here the spectral autocorrelation function
\begin{equation}\label{Fcorr}
F_{corr}(t-t') = \int_{E_{th}}^\infty  dE |\mu_{2E}|^2
e^{i\,\Delta_2\,(t-t')}
\end{equation}
is responsible for the field-induced decay from the state
$|2\rangle$ and subsequent re-pumping from the continuum, and the
source function
\begin{equation}\label{F0} F_0(t)  = \int_{E_{th}}^\infty dE\, b_E(0)
\,e^{i\Delta_2t}
\end{equation}
gives the phase-space envelope  \cite{zhenia-envelope} of the
initial wave packet of continuum states with near-threshold
energy.

Equation (\ref{b2}) can be further simplified by using the ''flat'',
or ''slowly varying continuum'' approximation (SVCA), which assumes
that $\Omega_E$ is constant within the range of continuum energies
affected by the laser pulses and that $E_{th}$ can be replaced with
$-\infty$. In the calculations below the spectral width of the
pulses is close to the temperature-defined energy spread in the
ensemble, so the non-uniform character of the FC factors for
continuum-bound transitions can in fact reveal itself. Nevertheless
the SVCA provides a good estimate: the Maxwell-Boltzmann statistical
weight of the lowest energy states, which poses the highest
challenge to the approximation, is small.

Within SVCA, $F_{corr}(\tau) = 2\pi\delta(\tau)$. This means that
the size of the Frank-Condon window in phase space is small
compared to the characteristic features of the continuum
wavefunction. The population transferred by the field from the
state $|2\rangle$ into continuum immediately leaves the FC window,
and cannot be re-pumped back. Taking the integral involving
$F_{corr}$ in Eq.(\ref{b2}) (note that the integral is taken on
the half, rather than whole, axis of time), yields the two coupled
equations describing the dynamics of the bound states in the
presence of the photoassociation pulses  \cite{SBbook,vardi97}:
\begin{eqnarray}\label{SVCA-1}
\dot{b_{1}}= i \,\Omega_1^*\, b_2 e^{-i\,\Delta_1\,t}
~~~~~~~~~~~~~~~~~~~~~~~~ \\\label{SVCA-2} %
\dot{b_{2}}= i\,
\Omega_1 \,b_1  e^{i\,\Delta_1\,t} - \G\,b_2 +
               i\, \Omega_E\,F_0(t)
\end{eqnarray}
where $\Gamma \equiv \Gamma_f + \pi |\Omega_E|^2$.

Appearance of Equations (\ref{SVCA-1},\ref{SVCA-2}), with the
initial (continuum) state eliminated, is different from the one
commonly used to describe bound-bound adiabatic transfer
\cite{stirap1,stirap2}. Diagonalization of Eqs.
(\ref{SVCA-1},\ref{SVCA-2}), unlike that of the conventional
equations for bound-bound passage, does not explicitly reveal the
''dark'' state connecting the initial and the target states of the
system. Instead, the continuum acts as a source of population for
the final state $|1\rangle$, with a very small transient
population in the intermediate state $|2\rangle$ during the
coherent "scooping" process. We choose this description versus the
ones which replace the continuum by a single, decaying
\cite{stirap2,kobrak-rice} or non-decaying \cite{mackie-PRA99},
level, since it allows to explicitly take into account the shape
of the initial continuum wave packet. This approach is necessary
when considering photoassociation with pulses, which coherently
pumps up by the field at $\omega_2$ a band of continuum levels of
the width $\Delta E \sim \Delta\omega_2$, creating a ''dark'' wave
packet on the background of the initial continuum state. Equations
\ref{RWA1-3}, \ref{SVCA-1} and \ref{SVCA-2} connect the shape of
the wave packet scooped from the continuum with that of the laser
pulses. For a given $F_0(t)$, representing the initial wave packet
of continuum states, one can find, in principle, optimal pulses
$\W_1(t)$ and $\W_E(t)$ that will photoassociate the entire
population contained in it \cite{ShSh-inprep}.
It has been demonstrated that a ``counterintuitive'' PAP sequence
of Gaussian pulses, analogous to a bound-bound STIRAP sequence,
can give rise to an almost complete population transfer in the
Na+Na$\rightarrow$Na$_2$ photoassociation \cite{vardi97}.

\begin{figure}
\centering
    \includegraphics[width=0.95\columnwidth]{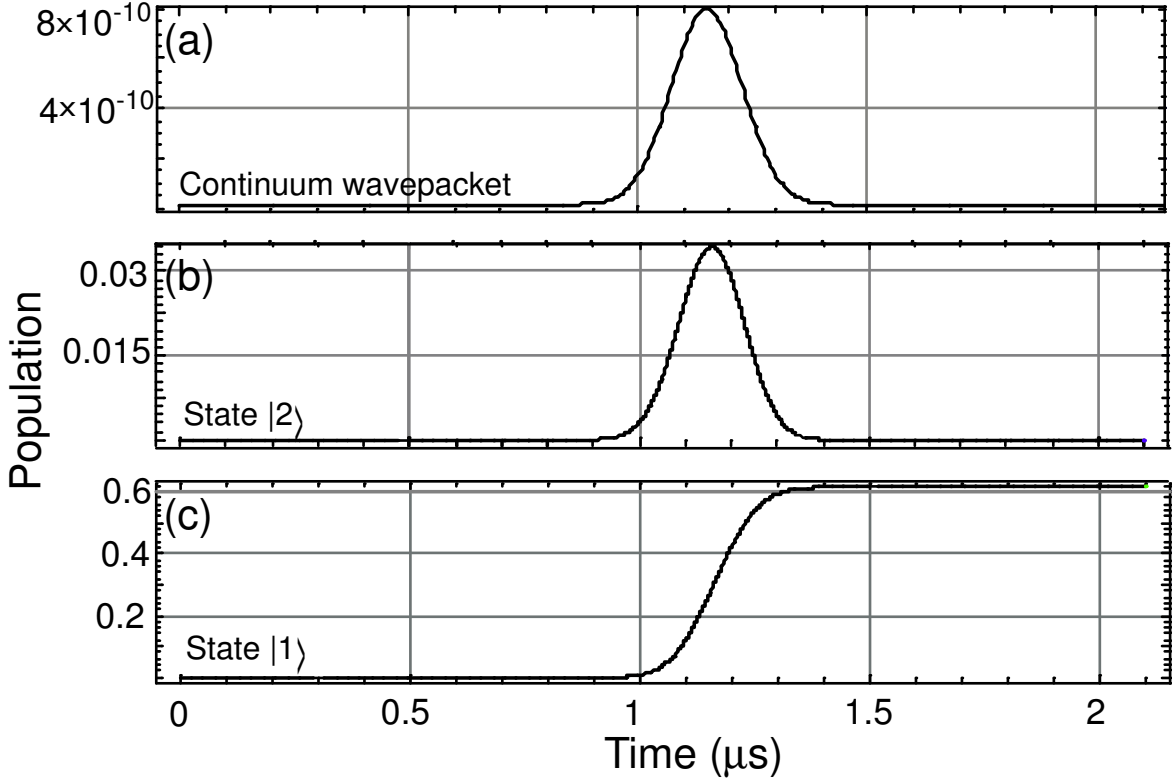}
    \caption{Photoassociation of a coherent wave packet.
(a)   Initial continuum probability distribution $|F_0(t)|^2$,
a.u.; (b) The population of the intermediate A/b state; (c) The
population of the X($v=4,J=0$) target state.}
\label{FigCoherentPA}
\end{figure}

Figure~\ref{FigCoherentPA} shows one simple example of
photoassociation of a coherent Gaussian wave packet in Rb, which
is the shortest in time allowed by the temperature spread $\delta
E$ (transform limited wave packet). Intuitively this is the most
classical scattering scenario allowed under the quantum law, where
two atomic wave packets of minimal spread collide at time $t_0$.
The initial state is given by the superposition of the energy
eigenstates
\begin{eqnarray}
b_E(t=0)=(\delta_E^2 \pi)^{-1/4}~\exp\left[-(E-E_0)^2/2\delta_E^2 +
                          i (E-E_0)t_0\right]
\label{coherentWP-initial}
\end{eqnarray}
with $E_0=100$ $\mu K$, $\delta_E = 70$ $\mu K$, $t_0=1150$ ns.
The pulse pair overlap time is chosen to match the coherence time
of an atom-pair during collision at a given temperature and $t_0$
corresponds to the maximal overlap between the $\W_1(t)$ and
$\W_E(t)$ pulses. The resulting continuum envelope $F_0(t)$ is
shown in Fig.~\ref{FigCoherentPA}(a). In this calculation, the
applied pair of laser pulses transfers the entire population of
the continuum wave packet to the X($v=4, J=0$) state (of energy $E
=-0.01823$ a.u.). The pair of pulses is composed of a $\sigma^-$
polarized 733 nm Stokes pulse, whose intensity is $7\times 10^3$
W/cm$^2$, operating in resonance with $\omega_{2,1}$ transition
frequency between the final $|1\rangle={\rm X}(v=4,J=0)$ state and
the intermediate $|2\rangle={\rm A/b}(v=133,J=1)$ state. The
second component is a $\sigma^+$ polarized pump pulse, operating
in resonance with $\omega_{2,E}$ transition frequency between the
intermediate ${\rm A/b}(v=133,J=1)$ state. and the X$(E,J=0)$
initial continuum state. The pump pulse, which is centered about
1063.4 nm, has intensity of $10^4$ W/cm$^2$. Both pulses are of
750 ns FWHM duration, with the pump delayed 600 ns relative to the
Stokes pulse (the ``counter-intuitive'' order).

The pulse durations are chosen so that their spectral widths will
roughly coincide with the initial continuum energy spread (which
determines $F_0(t)$). As shown below, this fact ensures that the
yield remains high even after thermal averaging at $T=100$ $\mu$K is
performed.

Assuming a $\sin^2(\alpha t)$ - shaped pulses and setting
$1/\Gamma_f=30$ ns, we numerically solve the Schr\"odinger
equation from which the continuum has been adiabatically
eliminated. As shown in Fig.~\ref{FigCoherentPA}, the final
population of the X($v=4,J=0$) state is 0.6. When $\Gamma_f$ is
set to zero, $P(E)$, the transfer probability per collision may
reach values as high as 0.9 for the same pulse configurations.

Once the X($v = 4,J=0$) state is populated, it is easy to find an
additional pair of pulses of variable intensities and center
frequencies that can execute the transfer to the desired X($v =
0,J=0$) state. An example of such a pair of pulses, applied right
after the end of the first pair, is shown in
Fig.~\ref{FigThreeways}. Both pulses are circularly polarized,
750 ns FWHM in duration, and have intensity of 10W/cm$^2$. The
transfer proceeds via an intermediate A/b vibrational state
$|3\rangle$ of energy $E_1= 0.03309$ a.u. The pump pulse of
wavelength $888$nm is followed after a delay of 600 ns by a Stokes
pulse of wavelength $869.6$nm. As we shall see, the population of
the X($v=4,J=0$) state is completely transferred by this pair to
the X($v=0,J=0$) state (state $|4\rangle$ in
Fig.~\ref{FigensemblePA}).

In order to estimate the fraction of atoms photoassociated per
pulse-pair we need to multiply $P(E)$, the photoassociation
probability per collision at energy $E$, by the number of
collisions experienced by a given atom during the pulses (this is
equivalent to averaging over all possible values of $t_0$). The
number of collisions during the pulses is calculated as follows:
At a given energy $E$, the velocity of a given atom is
$v=(2E/m)^{1\over2}$ and the distance traversed by it during a
pulse of $\tau_{laser}$ duration is $v\tau_{laser}$. The
cross-section for collision is $\pi b^2$ where $b$ is the impact
parameter, related to the $J$ partial wave angular momentum as
$b=(J+1/2)/p=(J+1/2)/(2mE)^{1\over 2}$. Hence, the number of
collisions experienced by the atom during the two pulses is
$N=\rho\pi b^2 v\tau_{laser}$ where $\rho$ is the density of
atoms. Putting all this together we have for $J=0$ that the
fraction of atoms photoassociated per pulse-pair is
\begin{equation}\label{estimate}
f(E)={P(E)\pi\rho(2E/m)^{1\over 2}\tau_{laser}\over 8mE}=
{P(E)\pi\rho\tau_{laser}\over 4m^{3/2}(2E)^{1\over 2}}~.
\end{equation}
When optimizing the photoassociation efficiency for an atomic
ensemble, $\tau_{laser}$ is not a free parameter: bandwidth of the
overlap of the pump and the Stokes pulses should match the energy
spread in the ensemble $kT$.  For $\tau_{laser}=750$ ns, atomic
density in an atomic trap of density $\rho=10^{11}$ cm$^{-3}$,
collision energy $E=100$ $\mu$K, reduced atomic mass of
$m=1823\times 85/2$ a.u., we have that $f\approx 4\times 10^{-7}$
per pulse-pair.


As an alternative to the above estimate, one can repeat the
calculations starting with an almost mono-energetic continuum wave
packet that is spread over the average distance between two atoms
experiencing an s-wave collision: $r_{st} = 1/(\pi\rho b^2)$. For
a collision at 100 $\mu$K energy, and the atomic density in the
trap equal to $\rho=10^{11} $cm$^{-3}$, we obtain
$r_{st}=4.21\times10^9$ a.u. From the uncertainty principle, one
must set the energy spread of such wave packets to $\Delta E =
\sqrt{E/2m\,r^2_{st}} =1.07\times10^{-17}$ a.u. in
Eq.(\ref{coherentWP-initial}).   Then $F_0(t)^2 =
3.8\times10^{-17}$ a.u. at the time when the pulses are applied.

\begin{figure}
\centering
    \includegraphics[width=0.76\columnwidth]{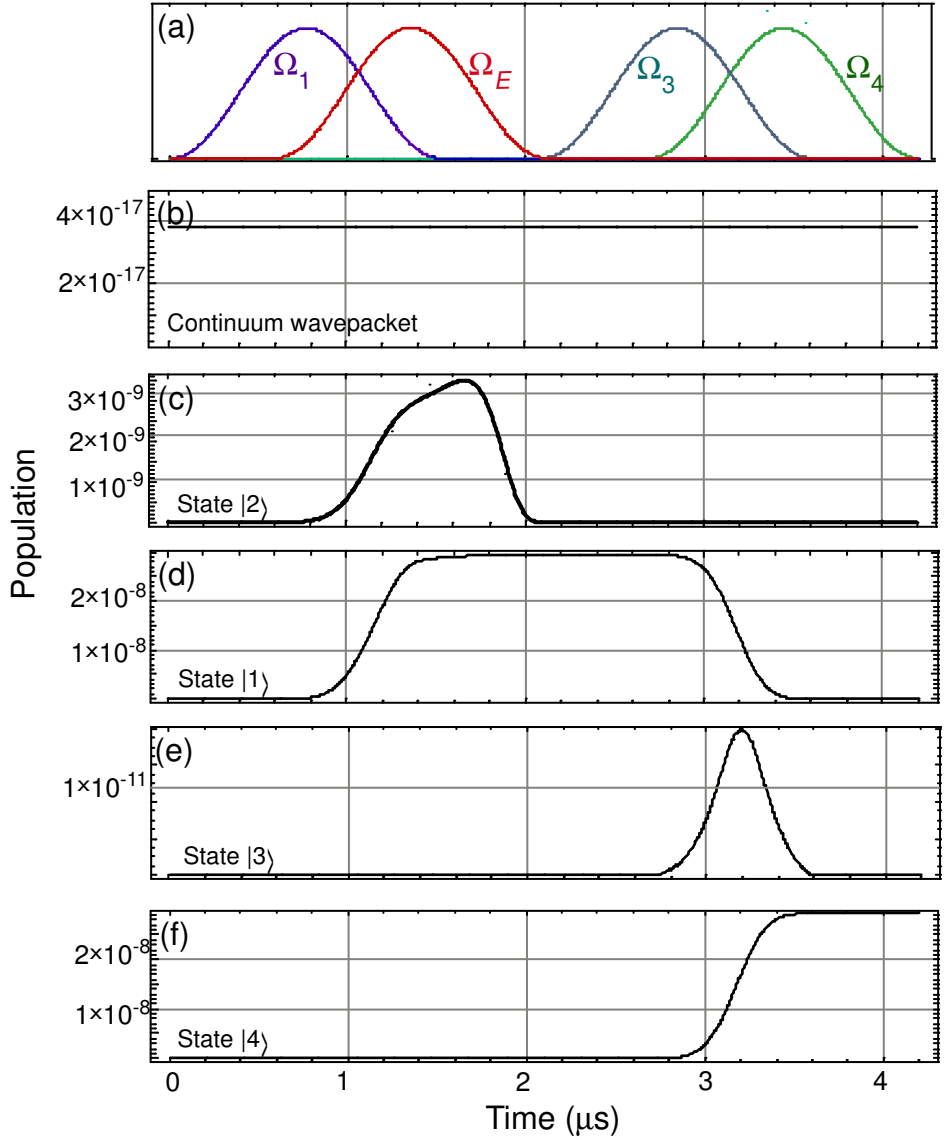}
    \caption{(Color online). Photoassociation of the atomic ensemble in a trap.
    (a): Envelopes of the four laser pulses, unscaled. (b): Continuum population
    $F_0(t)^2$;
(c)-(f): Bound state populations, weighted over the ensemble,
    for different bound states.
(c): Intermediate state $|2\rangle\equiv\,$A/b$(v=133,J=1)$,
$E=0.042848$ a.u; (d): State $|1\rangle\equiv\,$X$(v=4,J=0)$,
$E=-0.01823$ a.u.; (e): Intermediate state
$|3\rangle\equiv\,$A/b$(v=35,J=1)$, $E=0.03309$ a.u.;
     (f): State $|4\rangle\equiv\,$X$(v=0,J=0)$, $E=-0.0193$ a.u.  }
\label{FigensemblePA}
\end{figure}

Since the wave packets are almost mono-energetic, it is now
necessary to average the results of these calculations over the
energy $E$ of the Maxwell-Boltzmann thermal distribution:
\begin{equation}\label{averaging}
P_{total} = \frac{2}{\sqrt{\pi}(kT)^{3/2}}\int_{E_th}^\infty dE\,
P(E) \sqrt{E}\exp[-E/kT] .
\end{equation}
 For each energy
the bound-continuum matrix element was obtained according to the
data of Fig.~\ref{FigFCfits}, and the time dependent Shr\"odinger
equation solved in the SVCA. The SVCA-based elimination of the
continuum becomes less accurate as the collision energy tends to
zero, but since the density of the continuum states vanishes at
small energies, the contribution to the weighted signal from the
low-energy collisions is small \cite{wigneradjustment}.

Fig~\ref{FigensemblePA} shows the ensemble-averaged populations of
all the states involved in the scheme under the action of the two
pulse pairs described earlier. The final population of the ground
X state is equal to $2.13\times 10^{-7}$. This number is lower
than the one given by the simple estimate above, since in the
realistic calculation the photoassociation takes place when the
two pulses overlap, rather than during the full $\tau_{laser}$.
Thus one needs to repeat the pulse sequence about $5\times10^6$
times in order to photoassociate the entire ensemble, assuming
that all the collisions take place on the $\Xs$ potential surface
of the Rb$_2$ molecule. Approximately three quarters of the
collisions occur on the $1^3\Sigma_u^+$ potential surface, and
these are not affected in our arrangement due to the selection
rules for transitions at the chosen frequencies. Accounting for
the latter collisions, we multiply the above number by 4, and
obtain that $\sim 2\times10^7$ pulses are required for complete
photoassociation.

When we come to estimate the maximal rate at which the
photoassociation pulse sets can be sent into the cold atomic
cloud, it is necessary to appreciate the need to avoid excitation
and heating by subsequent pulse sets of the molecules already
created. In order to accumulate a large number of cold molecules,
the molecules must be removed from the beam path of the light
after every set of pulses before the next set of pulses can be
sent in. This requirement poses an inherent limit on the actual
repetition rate of a possible experiment, which depends on the
mechanism used to remove the molecules. The simplest possibilities
of thermal diffusion / gravitational fall are slow processes, and
would limit the repetition time to 1-10ms. Faster mechanisms may
require application of mechanical forces on the molecules (but not
the atoms) to push them out of the beam. An example for a pushing
mechanism can be a fast moving optical lattice, made up of two
counter propagating waves of slightly different frequencies.  {
Suppose that we deal with an elongated trap, $\sim$20$\mu$m in
diameter, and $\sim$200$\mu$m long. Let us send in a slowly moving
optical lattice, oriented parallel to the long dimension of the
trap, moving with the speed of 20 cm/s (thermal velocity at the
100$\mu$K temperature). The lattice with the period of 1$\mu$m is
designed to hold only molecules (without heating) but not atoms. A
single pulse sequence will place about 20 molecules into every
$1\times 1\times 200$ $\mu$m slot of the lattice overlapping with
the trap. The next similar pulse sequence can be applied 100
$\mu$s later, when the molecule-loaded lattice slots leave the
trap.
}

 {The inherent inefficiency of the photoassociation process in
the coherent regime is due to the fact that one has to operate
within the coherence time of the ensemble, hiding the created
molecules away before the next pulse sequence arrives. A rather
common route around this difficulty relies on employing
{spontaneous} decay from an excited {state} to the ground
electronic potential, which hides the population away form the
laser field. Let us come at some moment after the REAR
photoassociating pulse pair is over, {and transfer all the
population accumulated at the state X($v=4$) to one of the excited
vibrational states of the A/b potential (e.g. by adiabatic rapid
passage with a chirped pulse or by applying a $\pi$-pulse)}. Free
decay from the latter state will incoherently populate
ro-vibrational levels of the X surface according to their FC
overlaps with the decaying A/b state. Note that the decaying A/b
state can be chosen quite freely, since we populate it by a
transition from the convenient $v=4$ state of the X potential,
rather than from a continuum state.{We can therefore choose to
populate the excited state with the best possible overlap to our
target ground level ($v=0$).} The second (bound-bound excitation
and decay) stage in this photoassociation scheme does not suffer
from the poor continuum-bound FC overlap, and the ground-state
molecules, although vibrationally hot, can be produced at rater
high rates.}

\begin{figure}
\centering
    \includegraphics[width=0.95\columnwidth]{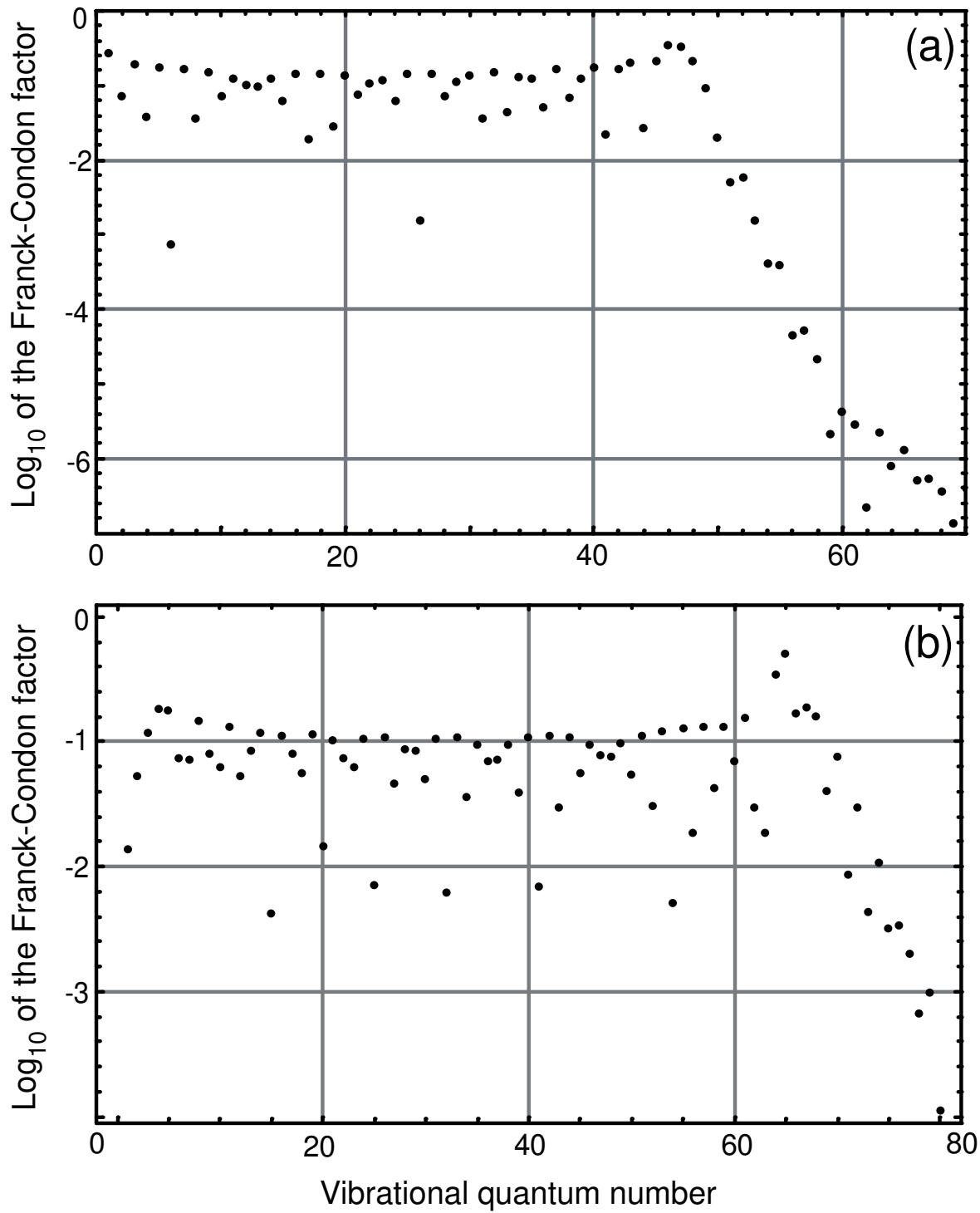}
    \caption{FC factors for the A/b$-$X bound-bound
transitions from the states A/b($v=51$) (a), and A/b($v=85$) (b)
in dependance on the vibrational quantum number of the X
potential.} \label{FigIncoherentFCs}
\end{figure}

{Fig.~\ref{FigIncoherentFCs}(a) shows the FC factors for
transitions between the state A/b($v=51$) ($E=0.0348$ a.u.) and
the vibrational states of the X potential. The FC factor for the
transition from the state X($v=4$) onto A/b($v=51$) is equal to
0.175. A picosecond-long $\pi$ pulse enforcing full transfer of
population between these levels would have the average intensity
of $7.4\times10^8$ W/cm$^2$. The FC factor for the transition from
the state A/b($v=51$) onto the state X($v=0$) is equal to $0.27$,
and so about 7.5\% of the population decaying from A/b($v=51$)
comes onto the ground vibrational state of the X potential. Let
us, again, assume that the molecules are created in a 200$\mu$m
long trap, and that the focus of the photoassociating laser is
5$\mu$m in diameter. A single PAP pulse sequence lasts,
approximately, 2$\mu$s, and creates about 400 molecules. Assuming
the two-stage pulse sequences to be repeated one after another,
and taking in account that 7.5\% of the created molecules are
accumulated at X$(v=0)$, we evaluate the rate of production of
molecules in the vibrational ground state as, approximately,
$1.5\times10^7$ per second.}

{{If we allow the molecules to accumulate incoherently in a
vibrational level that is not $v=0$ of the ground X potential, it
is possible to select an excited state that decays to a very
narrow set of ground vibrational levels}.
Fig.~\ref{FigIncoherentFCs}(b) shows the FC factors for
transitions between the state A/b($v=85$) ($E=0.0383$ a.u.) and
the vibrational states of the X potential. The FC factor for the
transition from the state X($v=4$) onto A/b($v=85$) is equal to
0.174, and so a picosecond-long pulse of the intensity
$\sim8\times10^8$ W/cm$^2$ can transfer all the population from
X($v=4$) onto A/b($v=85$).  The sharp maximum at the FC factor
corresponding to the transition between the states A/b($v=85$) and
X($v=64$) ($E=-0.0049$ a.u.) ensures that about 25\% of the
population decaying from the state A/b($v=85$) is accumulated at
the state X($v=64$), while the neighboring vibrational states are
barely populated. Indeed, an additional STIRAP implemented after
the photoassociation is completed, can transfer all the population
accumulated at X($v=64$) onto X($v=0$). }

{Our estimates are to be compared with the results on creation of
{ground state} molecules with CW lasers. In the experiment
 \cite{DeMille-PA05}, vibrationally cold ground state molecules
were created with the estimated rate of about 500 molecules per
second (since the molecules were not trapped, the actual detection
rate was lower). This value can be, in principle, increased; there
is a constraint on the effectiveness of the CW photoassociation
lying in the fact that it incoherently populates many highly
excited vibrational states, as well as the molecular continuum. In
the experiment  \cite{Nikolov-PRL00}, vibrationally hot ground
state molecules were created with a rate of $\sim 10^5$ molecules
per second for many vibrational levels. Although in this
experiment the molecules in the state $v=0$ were not detected, it
was concluded that their production rate is of the same order of
magnitude. The scheme discussed in the previous paragraph can be
viewed as an extension of the two-step schemes of Refs.
 \cite{DeMille-PA05,Nikolov-PRL00} to the case when the first step
is made with the help of the STIRAP-like PAP sequence. The high
efficiency appearing in our theoretical estimates comes from the
fact that PAP with correctly chosen parameters is assumed to waste
no molecules either in the continuum or} in the manifold of
molecular bound states.}

 {The estimate
(\ref{estimate}) trivially scales with the temperature and the
density of the atomic sample. In the case of photoassociation at
either BEC  \cite{BECs} or Mott insulator  \cite{Mott} conditions
one can hope to photo-associate the whole sample with a single
sequence of pulses.}

\begin{figure}
\centering
    \includegraphics[width=0.7\columnwidth]{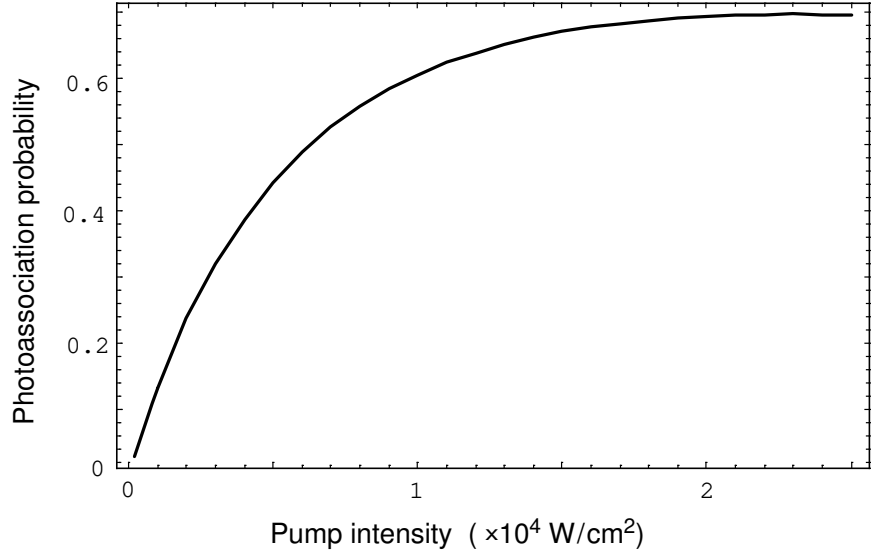}
    \caption{ Final population of the state X($v=4,J=0)$ in
photoassociation of a
      coherent wave packet in dependence on the pump pulse intensity.}
     \label{Fig-weakfield}
\end{figure}

Unlike bound-bound STIRAP, the continuum-bound REAR process
continues to work even if the laser intensity is not sufficient
for the complete population transfer. This can be seen from
Eq.(\ref{SVCA-2}). Apart from modifying the rate of decay of the
intermediate level, the pump Rabi frequency $\W_E$ enters the
Schroedinger equation only in combination with the initial
continuum envelope $F_0$. Reducing the pump Rabi frequency is
roughly equivalent to reducing the amplitude of the initial
scattering state, and does not influence the passage itself.

We repeated the calculation presented in Fig.~\ref{FigCoherentPA}
for a set of the pump intensities below and above $10^4$ W/cm$^2$.
The results are shown in Fig.~\ref{Fig-weakfield}.  As expected,
the photoassociation probability at low intensities roughly scales
as square root of the intensity. In each simulation the
probability of populating the intermediate A/b$(v=133,J=1)$ state
was much lower than the final photoassociation probability, and
the general time dependance of the bound state populations was
similar to that shown in Fig.~\ref{FigCoherentPA}.

Thus even in situations when the FC factors are too low for the
complete population transfer, a REAR photoassotiating pulse pair
allows to scoop from the continuum a part of its population, and
to transfer it, almost without losses, onto the bound molecular
state.

\section{Conclusions.}

We have demonstrated that one can implement AP in photoassociation
in a thermal ensemble of cold atoms. AP is best implemented when
the bandwidths of the laser pulses involved are slightly narrower
than the energy spread of the ensemble. In this way, the REAR
process works for any initial continuum energy.

The number of pulses needed to complete the photoassociation
depends only on the ratio between the coherence time and the
average time between s-wave collisions. Our estimate for the
number of pulses needed to photoassociate an entire ensemble of
100 $\mu$K Rb atoms at density of $10^{11}$ cm$^{-3}$ is
$2\times10^7$. The exact number can vary depending on whether one
photoassociates atoms colliding on the singlet (as we have done)
or the triplet potential surfaces.

The requirement that the photoassociating pulse sequence works the
same way for all the initial energies in the ensemble is generic
for any coherent photoassociation scheme. If simple laser pulses
are used, this means that the pulse durations must not exceed the
coherence time in the ensemble by much. The AP sequence considered
in this paper photoassociates almost {all} the continuum
population that passes through the FC region during the coherence
time window. Thus, for simple-structured pulses, the AP sequence
places the upper boundary on what is possible to achieve. Use of
more complex pulse shapes, such as the numerically optimized
shapes in Ref.  \cite{masnou2}, is limited in our scheme by the
requirement that the laser spectrum be smooth on the scale of the
ensemble thermal energy spread.

We conducted our simulations for singlet collisions of Rb$^{(85)}$
atoms. Due to the pole in the scattering length, the FC factors
are strongly enhanced, and the required intensities are quite low.
The scheme needs $10^4$ W/cm$^2$ intensity for the continuum-bound
1063.4 nm pump pulse; all the other intensities are adjustable,
and are below this number.

{Once the continuum state is transferred onto a vibrationally
excited bound state of the ground electronic surface, it is
relatively easy to employ another pulse sequence transferring the
population to the ground ro-vibrational state. This can be done
either as a coherent transition driven by a second AP pulse pair,
or by employing incoherent decay from an excited electronic state
of the molecule. In the first case, one creates vibrationally cold
molecules, which have to be removed from the laser focus before
the next photoassociating pulse pair arrives. In the second case,
vibrationally hot molecules in the ground electronic state can be
accumulated in the laser focus.}

{In an analogy with weak-field coherent control (CC)
 \cite{SBbook}, interference of quantum pathways for the system
following several dark states can either enhance or suppress the
probability of photoassociation. Combining the vast possibilities
of coherent control with the ability to adiabatically transfer
100\% of quantum population into the desired state  \cite{stirap2}
would give a powerful tool for quantum engineering. Progress in
this direction has been made in controlling bound-bound population
transfer  \cite{ioannis-review}.}

{We have numerically investigated the dark state interference for
free-bound transitions, for two basic configurations involving an
initial continuum -- the ``double-$\Lambda$'' (PAP via two
different intermediate states) and the ``tripod'' (PAP into a
superposition of two final states) linkages. In both cases the
branching ratios and the effectiveness of the population transfer
for the continuum-bound processes coincided with those expected
from the bound-bound multi-pathway STIRAP
 \cite{stirap2,ioannis-review}. Coherently Controlled Adiabatic
Passage, if implemented in photoassociation, can provide the means
to implement a number of exciting schemes based on the
continuum-bound pathway interference. As an example, it is
possible to measure the multi-channel continuum state in the trap
by photoassociating a desired multi-channel wave packet (such as
either even or odd superposition of single-channel components,
either even or odd temporal structure, etc.) at a given moment and
leaving aside all orthogonal waveforms  \cite{ShSh-inprep}. Some
work in this direction will be presented elsewhere. }

\section{Acknowledgments.}
The authors thank S. Lunell and D. Edvardsson for sharing the
output data of their calculation  \cite{edvadsson-03}. We
acknowledge fruitful discussions with M. Stowe, T. Zelevinsky, K.
Madison, and D. Jones. E. Shapiro is pleased to thank I.
Thanopulos for discussions and numerical advice. The work at JILA
is funded by NSF and NIST. A. Pe'er thanks the Fulbright
foundation for financial support.



\begin{thebibliography}{}
\bibitem{wigneradjustment}
It turns out that adjusting the matrix elements according to the
Wigner $E^{1/4}$ law rather than the resonance law would only
insignificantly decrease the final temperature-weighted population
of the ground state.
\end{thebibliography}
\end{document}